\documentclass[aps,prl,twocolumn,groupedaddress,showpacs,showkeys,amsmath]{revtex4-1}
\usepackage[dvips]{graphicx}
\usepackage{dcolumn}
\usepackage{bm}
\usepackage{color}
\usepackage{longtable}

\begin{document}

\title{Nature of the Metallization Transition in Solid Hydrogen} 

\author{Sam Azadi}

\affiliation{Department of Physics, Imperial College London, Thomas Young Centre and 
London Centre for Nanotechnology,  London SW7 2AZ, United Kingdom}

\email{s.azadi@ucl.ac.uk}

\author{N.\ D.\ Drummond}

\affiliation{Department of Physics, Lancaster University, Lancaster LA1 4YB,
  United Kingdom}

\author{W.\ M.\ C.\ Foulkes}

\affiliation{Department of Physics, Imperial College, Exhibition Road,
London SW7 2AZ, United Kingdom}

\date{\today}

\begin{abstract}
   We present an accurate study of the static-nucleus electronic energy band gap of
  solid molecular hydrogen at high pressure.  The excitonic and
  quasiparticle gaps of the $C2/c$, $Pc$, $Pbcn$, and $P6_3/m$
  structures at pressures of 250, 300, and 350~GPa are calculated using
  the fixed-node diffusion quantum Monte Carlo (DMC) method.
  The difference between the mean-field and many-body band gaps at the same
  density is found to be almost independent of system size and can therefore
  be applied as a scissor correction to the mean-field gap of an infinite
  system to obtain an estimate of the many-body gap in the thermodynamic
  limit. By comparing our static-nucleus DMC energy gaps with available
  experimental results, we demonstrate the important role played by nuclear
  quantum effects in the electronic structure of solid hydrogen.
  Our DMC results suggest that the metallization of high-pressure
 solid hydrogen occurs via a structural phase transition rather than band gap closure.  
\end{abstract}
\maketitle

Determining the metalization pressure of solid hydrogen is one of the
great challenges of high-pressure physics.  Since 1935, when it was
predicted that molecular solid hydrogen would become a metallic atomic
crystal at 25~GPa \cite{1935}, compressed hydrogen has been studied
intensively. Additional interest arises from the possible existence of
room-temperature superconductivity \cite{Ashcroft}, a metallic liquid
ground state \cite{Bonev}, and the relevance of solid hydrogen to
astrophysics \cite{Hemley,Ginzburg}.

Early spectroscopic measurements at low temperature suggested the
existence of three solid-hydrogen phases \cite{Hemley}.  Phase I, which is
stable up to
110~GPa, is a molecular solid composed of quantum rotors arranged in a
hexagonal close-packed structure. Changes in the low-frequency regions
of the Raman and infrared spectra imply the existence of phase II,
also known as the broken-symmetry phase, above 110~GPa.  The appearance
of phase III at 150~GPa is accompanied by a large discontinuity in the
Raman spectrum and a strong rise in the spectral weight of molecular
vibrons.  Phase IV, characterized by the two vibrons in its Raman
spectrum, was discovered at 300~K and pressures above 230~GPa
\cite{Eremets, Howie, Howie2}. Another new phase has been claimed to
exist at pressures above 200~GPa and higher temperatures (for example,
480~K at 255~GPa) \cite{Howie3}. This phase is thought to meet phases I
and IV at a triple point, near which hydrogen retains its molecular
character. The most recent experimental results \cite{Simpson} indicate
that H$_2$ and hydrogen deuteride at 300~K and pressures greater
than 325~GPa transform to a new phase V, characterized by substantial
weakening of the vibrational Raman activity. Other features include a
change in the pressure dependence of the fundamental vibrational
frequency and the partial loss of the low-frequency excitations.

Although it is very difficult to reach the hydrostatic pressure of more
than 400~GPa at which hydrogen is normally expected to metalize, some
experimental results have been interpreted as indicating metalization at
room temperature below 300~GPa \cite{Eremets}. However, other
experiments show no evidence of the optical conductivity expected of a
metal at any temperature up to the highest pressures explored
\cite{zha}.  Experimentally, it remains unclear whether or not the
molecular phases III and IV are metallic, although it has been suggested
that phase V may be non-molecular (atomic) \cite{Simpson}.  Metalization
is believed to occur either via the dissociation of hydrogen molecules
and a structural transformation to an atomic metallic phase
\cite{samprl,Eremets}, or via band-gap closure within the molecular
phase \cite{MStadele,KAJohnson}.  In this work we investigate the latter
possibility using advanced computational electronic structure methods.

Structures of crystalline materials are normally determined by X-ray or
neutron diffraction methods. These techniques are very challenging for
low-atomic-number elements such as hydrogen \cite{Dias}.
Fortunately optical phonon modes disappear, appear, or
experience sudden shifts in frequency when the crystal structure
changes. It is therefore possible to identify the transitions between
phases using optical methods.

The electronic structures of the solid molecular phases have mainly been
investigated using computational methods based on density functional
theory (DFT)
\cite{Pickard,Pickard2,Goncharov,Magdau,Naumov,Morales2013,Morales2014,JETP,singh,dft-fail}
and the quasiparticle (QP) approach within the $GW$ approximation
\cite{NJP,Johnson}.  Although DFT-based methods
can be used to search for candidate low-energy crystal
structures and to calculate their vibrational properties,
their inadequacies are more apparent in the case of band-gap
calculations \cite{JPPerdew}. To obtain precise gaps, it is vital to go
beyond mean-field-like methods and solve the many-electron
Schr\"{o}dinger equation directly.  In this work, we employ the
fixed-node diffusion quantum Monte Carlo (DMC) method to calculate
excitonic and QP band gaps of cold dense hydrogen as functions of
pressure.

Fixed-node DMC is the most accurate known method for evaluating the
total energies of continuum systems of more than
a few tens of interacting quantum particles 
\cite{Matthew1,ElahehPRL,Kolorenc,samjcp15,sambenz}. 
Recently, it has been indicated that DMC can provide an accurate description of
the phase diagram of solid molecular hydrogen \cite{Neil15}.  Although
the DMC method was originally designed to study ground states, it is
also capable of providing accurate information about excited states in
atoms, molecules, and crystals \cite{mitas,will,towler,Neil}.  DMC
calculations of excitations in crystals remain challenging because of
a $1/N$ effect: the fractional change in the total energy due to the
presence of a one- or two-particle excitation is inversely proportional
to the number of electrons in the simulation cell. Since large
simulation cells are required to provide an accurate description of the
infinite solid, high-precision calculations are necessary.

The main input to any {\it ab initio} calculation is the structure of
the system under study, which in this case is unknown. Hence there is no
option but to use structures predicted by mean-field methods such as
DFT\@. It is now generally accepted that DFT results for high-pressure
hydrogen depend on the choice of exchange-correlation functional
\cite{dft-fail,Morales2013,Morales2014}. This frustrating limitation may
be the main cause of the contradictions \cite{JMcMinis,JMMcMahon}
between existing computational results.

In the present work we use the DMC method to carry out a comprehensive
study of the pressure dependence of the energy band gap of solid
hydrogen at high pressure. The definitive static-nucleus many-body band
gap data we provide can be used to correct results obtained using less
accurate methods. The corrections required are approximately independent
of lattice vibrations and temperature.

We considered the $C2/c$, $Pc$, $Pbcn$, and $P6_3/m$ molecular
structures of solid hydrogen at pressures of 250, 300, and 350 GPa.  According to
{\it ab initio} calculations, the $C2/c$ and $Pc$ structures are the
most favorable candidates for phases III and IV, respectively
\cite{Pickard2,Neil15}.
The $C2/c$ and $Pc$ crystals have weakly-bonded graphenelike layers
\cite{Pickard2}, while the $Pbcn$ structure includes two different
layers of graphenelike three-molecule rings with elongated H$_2$
molecules and unbound H$_2$ molecules \cite{Pickard,Pickard2}. The $P6_3/m$
structure may also be viewed as layered but is not graphenelike: three
quarters of the H$_2$ molecules lie flat in the plane and one quarter
lie perpendicular to the plane. The interplane bonding is relatively
strong and the centers of the molecules fall on a slightly distorted
hexagonal close-packed lattice \cite{Pickard}.
The structures were fully relaxed using DFT at
fixed pressure, and the relaxed structures were used in the DMC
simulations. The details of our DFT calculations 
are provided in the Supplemental Material \cite{supp}.

The QP energy gap is defined as
\begin{equation}
\label{eq_qp}
\Delta_{\rm qp}= E_{N+1}+E_{N-1}-2E_0 ,
\end{equation}
where $E_0$ is the ground-state energy of a system of $N$ electrons and
$E_{N+1}$ ($E_{N-1}$) is the many-body total
energy of the system after an electron has been added to
(removed from) the system.
Our calculations of $\Delta_{\rm qp}$ are
performed at the $\Gamma$ point of the supercell Brillouin zone,
equivalent to a mesh of $k$-points including $\Gamma$ in the primitive
Brillouin zone. 
We calculate a vertical QP energy gap, assuming that the ground- and
excited-state structures are the same. The difference between the vertical
and adiabatic QP gaps is expected to be small \cite{Elaheh2}.
We create excitonic states by promoting an electron from a valence-band
orbital into a conduction-band orbital with the same Bloch
wavevector. The excitonic absorption gap is
\begin{equation}
\label{eq1}
 \Delta_{\rm exc} = E^{\prime} - E_{0},
\end{equation}
where $E^{\prime}$ is the total energy of the excitonic state. Again we
work at the $\Gamma$ point of the supercell Brillouin zone.  In the
ground-state geometry, the singlet excitonic gap is equivalent to the
vertical optical absorption gap \cite{Elaheh2}.  Our DMC calculations
used Slater-Jastrow trial wave functions as implemented in the
\textsc{casino} quantum Monte Carlo (QMC) code.  Further details of our
simulations and tables of the DMC total energies and band gaps as
functions of system size are given in the Supplemental Material
\cite{supp}.

We find that the singlet and triplet exciton binding energies in
high-pressure solid hydrogen are smaller than 0.1~eV and cannot be
resolved above the statistical and finite-size errors in our DMC
results.  Many-body perturbation theory calculations of the excitonic
gap of the $Cmca$-$12$ structure showed that the exciton binding energy
decreases with increasing pressure from 66~meV at 100~GPa to 12~meV at
200~GPa \cite{Dvorak}.  Accurate DMC calculations of the exciton binding
energy would therefore require an unattainable precision of better than
10~meV in the total energy of the simulation cell. Therefore, in the
rest of this paper, we do not attempt to distinguish the excitonic band
gap from the QP band gap..

The simplest possible antisymmetric many-electron trial wave function is
a Slater determinant of Hartree-Fock or DFT orbitals. Multiplying the
Slater determinant by a Jastrow factor helps to keep electrons away from
each other and significantly lowers the energy expectation value
calculated in a variational quantum Monte Carlo (VMC) simulation, but
does not change the nodal surface (i.e., the surface on which the
many-body wave function $\Psi(\bm{r}_1,\bm{r}_2,\ldots,\bm{r}_N)$ is
zero) or the fixed-node DMC energy. Introducing a backflow (BF)
transformation \cite{Lopez}, which can be viewed as a leading-order
improvement to the Slater-Jastrow form \cite{Holzman,Kwon}, changes the
nodes and thus lowers the DMC energy. Here we systematically investigate
the influence of BF on the fixed-node DMC results for solid hydrogen.
We also address the question of how the choice of wave function affects
VMC and DMC results.

Band gaps calculated using Hartree-Fock theory, which neglects
electron-electron correlation, are generally much too large. DMC
calculations using Slater-Jastrow trial wave functions retrieve a high
percentage of the correlation energy and produce gaps closer to
experimental values. It is unsurprising that improving the DMC
description of electronic correlation by adding a BF transformation
further lowers the calculated DMC gap. Using BF trial wave functions
decreases the calculated QP and excitonic gaps of the $C2/c$ structure
by 0.5(1) and 0.2(1)~eV, respectively, bringing them within error bars
of each other.  Although the inclusion of BF considerably improves the
DMC results, the computational cost is high. One of the most expensive
operations in any DMC code is the evaluation of the orbitals and their
first two derivatives, and the evaluation of the collective BF
coordinates makes this even slower, because every element of the Slater
matrix must be updated every time a single electron is moved.  For this
reason we did not utilize BF wave functions for the other structures at
different pressures.

To obtain DMC band gaps in the thermodynamic (infinite supercell) limit
we introduce a {\it scissor operator} $\delta_{\rm sci}(N)$, defined as
the difference between the DMC and DFT band gaps of a given supercell at
a given density: $\delta_{\rm sci} (N) = \Delta^{\rm DMC} (N) -
\Delta^{\rm DFT} (N)$, where $\Delta^{\rm DMC} (N)$ and $\Delta^{\rm
  DFT} (N)$ are DMC and DFT band gaps for a simulation cell containing
$N$ atoms.  Similar methods have been employed successfully for silicon
and germanium \cite{ZHLevine}.  The magnitude of the scissor correction
depends on the crystal structure and the applied pressure.  The values
of $\delta_{\rm sci}(N)$ for all of the structures and supercells
studied at pressures of 250, 300, and 350~GPa are given in the
Supplemental Material \cite{supp}. To within our statistical error, we
found that $\delta_{\rm sci}(N)$ is a constant [within $\pm 0.1(1)$ eV]
for $N \geq $200, independent of system size $N$. The DMC band gap at
infinite system size limit is therefore given by $\Delta^{\rm DMC}
(N\rightarrow \infty) = \Delta^{\rm DFT} (N\rightarrow \infty) +
\delta_{\rm sci}$.

Table \ref{static_gaps} shows the static-nucleus (Born-Oppenheimer) DMC
band gaps of the $C2/c$, $Pc$, $Pbcn$, and $P6_3/m$ structures at
pressures of 250, 300, and 350~GPa. The band gaps of the $C2/c$ and $Pc$
structures are similar, as are those of the $Pbcn$ and $P6_3/m$
structures.  The $P6_3/m$ band gaps are slightly greater than those of
the other structures studied.  A linear extrapolation suggests that the
band gaps of the $C2/c$, $Pc$, $Pbcn$, and $P6_3/m$ structures vanish at
pressures of 464(5), 421(6), 442(5), and 473(4)~GPa.  DMC calculations
of the phase diagram predict that the static molecular-to-atomic phase
transition also occurs in the pressure range 415--475~GPa \cite{samprl}.

\begin{table}[ht]
\caption{\label{static_gaps} DMC band gaps for different high-pressure
  solid molecular hydrogen structures at pressures of 250, 300, and 
`  350~GPa.}
\begin{tabular}{l c c c }
\hline\hline
 & \multicolumn{3}{c}{$\Delta^{\rm DMC}$ (eV)} \\

\raisebox{2ex}[0pt]{Structure~~} & 250~GPa~~  & 300~GPa~~ & 350~GPa~~  \\

\hline

$C2/c$  & 3.0(2)      & 2.3(2)       & 1.6(2) \\
$Pc$    & 3.2(2)      & 2.4(2)       & 1.3(2) \\
$Pbcn$  & 3.6(2)      & 2.8(2)       & 1.7(2)  \\
$P6_3/m$& 3.6(2)      & 2.8(2)       & 2.0(2)  \\
\hline\hline
\end{tabular}
\end{table}
 
\begin{figure}
\includegraphics[width=0.5\textwidth]{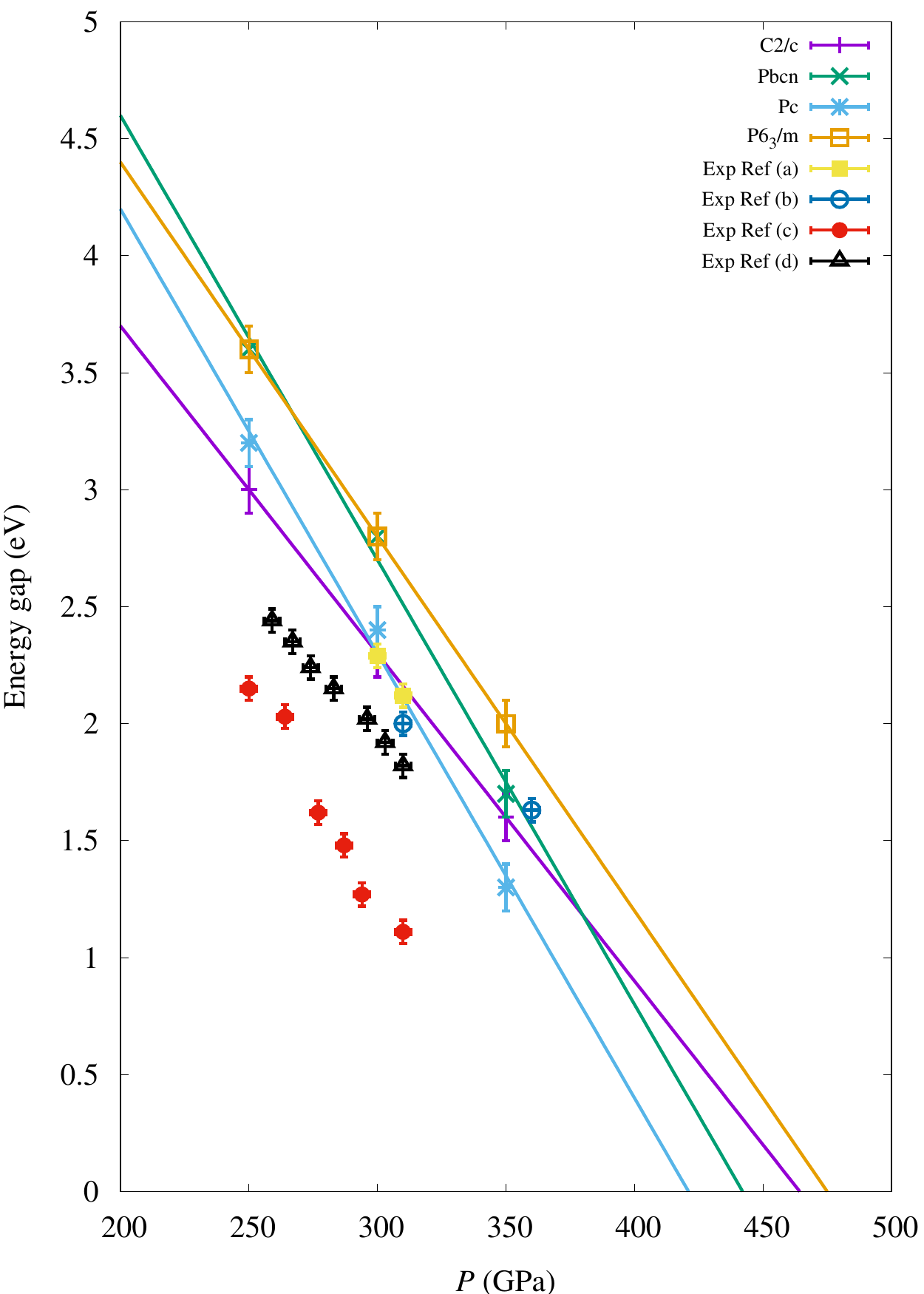}
\caption{\label{ground_gap} DMC energy gaps for the $C2/c$, $Pc$, $Pbcn$, 
  and $P6_3/m$ structures against pressure $P$. 
  References~(a) \cite{Loubeyre}, 
  (b) \cite{zha}, (c) \cite{Goncharov}, and (d) \cite{Howie} are energy gaps 
  at different $P$ ($\pm 3$ GPa) reported by experiments.} 
\end{figure}

Figure \ref{ground_gap} compares the pressure dependence of the
static-nucleus DMC band gaps of the $C2/c$, $Pbcn$, $Pc$, and $P6_3/m$
structures of hydrogen with experimental data. The DMC energy gaps
for the $Pc$ and $C2/c$ structures at 300~GPa are close to the
absorption-edge measurements for hydrogen at 100~K and above 300~GPa
reported in Ref.~\onlinecite{Loubeyre}.  These authors predicted that,
at low temperatures, metallic hydrogen should be observed at about
450~GPa, when the electronic band gap closes.  
The $Pbcn$ and $P6_3/m$ energy gaps are larger than the
experimentally measured gap over the entire pressure range studied.
Figure \ref{ground_gap} illustrates that 
there is a substantial disagreement between experimental gap measurements.

It is well known \cite{samprl,Neil15} that nuclear quantum effects
(NQEs) are significant in hydrogen-rich systems and affect the phase
transitions of high-pressure solid hydrogen. DFT-based path-integral
molecular dynamics (PIMD) simulations \cite{Morales2013} indicate that
the influence of NQEs on the band gap is strongly dependent on the
choice of exchange-correlation functional. PIMD results at $T=$ 200~K
obtained using the Perdew-Burke-Ernzerhof (PBE) \cite{PBE} functional
predict that the band gaps of the $C2/c$ and $Pbcn$ structures close
below 250~GPa \cite{Morales2013}, in disagreement with experiment.  PIMD
simulations employing the Heyd-Scuseria-Ernzerhof (HSE) \cite{HSE}
functional are not significantly better, although using a van der Waals
functional leads to an improvement \cite{Morales2013}. These results are
surprising because DFT calculations using the hybrid HSE functional
normally yield much better ground-state band gaps than calculations
using the semi-local PBE functional \cite{Paier}.

Assuming the validity of the Born-Oppenheimer approximation, the full
electron-nuclear wave function $\Psi({\bf R}, {\bf d})$ may be
approximated as $\Phi({\bf R} | {\bf d}) \chi({\bf d})$, where
$\Phi({\bf R} | {\bf d})$ is a function of the positions ${\bf R} =
({\bf r}_1,{\bf r}_2,\ldots,{\bf r}_N)$ of the $N$ electrons in the
supercell at fixed nuclear positions ${\bf d}$, and $\chi({\bf d})$ is
the nuclear wave function. The band structure as calculated using PIMD
is an average of the band structures corresponding to the electronic
wave functions $\Phi({\bf R} | {\bf d})$, weighted according to the
nuclear probability density $|\chi({\bf d})|^2$. Since each HSE band gap
is likely to be better (wider) than the corresponding PBE band gap, the
finite-temperature HSE-PIMD gap ought to be better than the PBE-PIMD
gap.  The observation that both PIMD gaps are poor suggests, therefore,
that \emph{both} functionals produce inaccurate nuclear probability
densities $|\chi({\bf d})|^2$. This problem is consistent with other
observed failures of DFT for high-pressure hydrogen \cite{dft-fail}.
Understanding the influence of NQEs and temperature on the band gap of
solid hydrogen is a challenging problem that may require going beyond
DFT-based methods. We do not address this problem here, but a comparison
of our static-nucleus DMC band gaps with experimental results can yield
estimates of NQEs.

It is not straightforward to measure the band gap at pressures greater
than 300~GPa, but the experimental results shown in
Fig.~\ref{ground_gap} suggest that solid hydrogen remains an insulator
up to 350~GPa or more. The $C2/c$ and $Pc$ structures are currently
considered \cite{Neil15} the most likely candidates for phases III and
IV, respectively, and the $Pbcn$ and $P6_3/m$ structures have higher
band gaps than these. Despite the inevitable band-gap reduction due to
NQEs, it is reasonable to assume that all of the structures considered
in this paper have nonzero band gaps at 300~GPa and 300~K\@ and beyond.
The estimated molecular-to-atomic transition pressure, calculated using
static-nucleus DMC calculations together with DFT anharmonic vibrational
corrections, is about 374 GPa \cite{samprl}.  According to Fig.~\ref{ground_gap},
the vibrational renormalisation of the gap of C2/c (the structure believed
to correspond to phase III) would have to be an implausibly large $-1.3(2)$ eV
if the gap is to have closed at 374 GPa.  Hence our results suggest that the
metallisation of hydrogen does not occur via closure of the band gap of
the molecular phases, but rather by a structural phase transition to an atomic phase

The main effect of quantum and thermal vibrations is to increase the
intermolecular interactions and weaken the intramolecular bonding.
Bearing in mind the symmetries and geometries of the crystals studied,
we would expect the NQE-induced band-gap reduction to be larger in the
layered $C2/c$, $Pc$, and $Pbcn$ structures than in the $P6_3/m$
structure.  This suggestion is consistent with the high structural
flexibility of phase IV observed in {\it ab initio} variable-cell MD
simulations \cite{HLiu} at pressures of 250--350~GPa and temperatures of
300--500~K. Protons in the graphenelike layers were seen to transfer
readily to neighboring molecular sites via a simultaneous rotation of
three-molecule rings.

As illustrated in Fig.\ \ref{ground_gap}, there are experimental
band-gap results up to 350~GPa. The optical transmission spectrum of
phase IV shows an overall increase of absorption and a reduction of the
band gap to 1.8~eV at 315~GPa \cite{Howie}. According to these results,
solid hydrogen at and below room temperature should metalize at
pressures above 350~GPa, in good agreement with our static-nucleus
results. Assuming that the $Pc$ structure is the best candidate for
phase IV, as has been reported recently \cite{Neil15}, the calculated
DMC band gap at 300 GPa is 0.6(2)~eV larger than the experimental gap
\cite{Howie}.  The difference is similar to the zero-point
renormalization of the diamond band gap at ambient conditions, which was
found to be as large as 0.6~eV \cite{Giustino, Marini}, but the atomic
mass of carbon is twelve times that of hydrogen and we would expect a
larger band-gap reduction here.  Other experimental results
\cite{Goncharov} report an energy gap of 1.2~eV for high pressure
hydrogen at 300~K and pressures around 300~GPa. This would imply a NQE
band-gap reduction of 1.2(2)~eV, which we believe to be more plausible.
Bearing in mind the expected NQE, our static-nucleus
DMC gaps are more consistent with the experimental results reported in
Ref.~\onlinecite{Goncharov} than with those reported in
Refs.~\onlinecite{Loubeyre, zha, Howie}. 

In summary, we have performed DMC calculations of the QP and excitonic
energy band gaps of solid molecular hydrogen at high pressure. We find
that the exciton binding energy is smaller than 100~meV/atom and that
our DMC QP and excitonic band gaps are within error bars of one
another. We have systematically investigated the energy reductions
obtained by introducing a better description of electronic correlation
into our VMC and DMC trial wave functions. Using a highly-correlated BF
wave function reduces the DMC band gap and significantly improves the
ground-state DMC energy by decreasing the FN errors. NQEs reduce the
band gap significantly, but a comparison of our DMC band-gap results
with experiments suggest nevertheless that the metalization of solid
hydrogen at and below room temperature occurs via a structural
transformation rather than band-gap closure. 
Assuming the existence of high pressure cold
liquid state \cite{Bonev,Eremets}, our results indicate that the
molecular phase is the only solid phase of hydrogen.  It is plausible
that melting curve separates the solid and liquid phases, the
insulating and metallic phases, and also the molecular and atomic
phases.

This work was supported by the UK Engineering and Physical Science
Research Council under grant EP/K038141/1, by the Thomas Young
Centre under grant TYC-101, and by PRACE-3IP project FP7
RI-312763. Computing facilities were provided by ARCHER, the UK National
Supercomputing Service, and by the Imperial College London High
Performance Computing Centre.

\end{document}